\documentclass{WileyMSP-template}

\usepackage{bm,amsmath,amsfonts,color}

\def\be{\begin{eqnarray}}
\def\ee{\end{eqnarray}}
\def\ii{{\mathrm i}}
\newcommand{\Imag}{\mathrm{Im} \, }
\newcommand{\Real}{\mathrm{Re} \, }

\def\r{{\bm r}}

\def\II{{\overleftrightarrow{{\bf I}}}}
\def\GG{{\overleftrightarrow{{\bf G}}}}
\def\GGL{{\overleftrightarrow{{\mathbb{G} }}}}
\def\alphagg{{\overleftrightarrow{{\bm{\alpha}}}}}

\begin{document}

\pagestyle{fancy}

\title{Tailoring accidental double bound states in the continuum in all-dielectric metasurfaces}

\maketitle


\author{Diego R. Abujetas*}
\author{Jorge Olmos-Trigo}
\author{Jos\'e A. S\'anchez-Gil*}


\dedication{}

\begin{affiliations}
Dr. Diego R. Abujetas\\
Physics Department, Fribourg University, Chemin de Musée 3, 1700 Fribourg Switzerland \\
Email Address: diego.romeroabujetas@unifr.ch

Dr. Jorge Olmos-Trigo\\
Donostia International Physics Center (DIPC), 20018 Donostia-San Sebastian, Spain 

Dr. Jos\'e A. S\'anchez-Gil\\
Instituto de Estructura de la Materia (IEM-CSIC), Consejo Superior de Investigaciones Cient\'{\i}ficas, Serrano 121, 28006 Madrid, Spain\\
Email Address: j.sanchez@csic.es

\end{affiliations}


\keywords{accidental bound states in the continuum, all-dielectric metasurfaces, Mie-resonant nanostructures}


\begin{abstract}

Bound states in the continuum (BICs) have been thoroughly investigated due to their formally divergent Q-factor, especially those emerging  in all-dielectric, nanostructured metasurfaces from symmetry protection at the $\Gamma$ point (in-plane wavevector $k_{||}=0$). Less attention has been paid to accidental BICs that may appear at any other $k_{||}\not =0$ in the band structure of supported modes, being in turn difficult to predict. Here we make use of a coupled electric/magnetic dipole model to determine analytical conditions for the emergence of accidental BICs, valid for any planar array of meta-atoms that can be described by dipolar resonances, which is the case of many nanostructures in the optical domain. This is explored for all-dielectric nanospheres through explicit analytical conditions that allow us in turn to predict accidental BIC positions in the parameter space $(\omega,\bf{k_{||}}$). Finally, such conditions are exploited to determine not only single, but also double (for both linear polarizations) accidental BICs occurring at the same position in the dispersion relation $\omega-\bf{k_{||}}$  for realistic semiconductor nanodisk meta-atoms. This might pave the way to a variety of BIC-enhanced light-matter interaction phenomena at the nanoscale such as lasing or non-linear conversion, that benefit from emerging at wavevectors away from the  $\Gamma$ point (off-normal incidence) overlapping for both linear polarizations.

\end{abstract}


\section{Introduction}

Planar arrays of particles, often called metasurfaces, are versatile thin platforms designed to control the properties of light at both the far- and the near- field, that have opened up new opportunities to tailor light-matter interaction by customizing the electromagnetic field~\cite{Holloway2012a,Glybovski2016,Haltout2016,Li2017c,Genevet2017,Neshev2018,Qiao2018,Baur2018,Sun2019a,Kupriianov2019,Paniagua-Dominguez2019,Staude2019,Sain2019a,Pertsch2020,Tsilipakos2020,Cuartero-Gonzalez2020}. The shape, position, and material properties of each particle in the array can be freely designed in order to produce desired wavefronts (known as Huygens metasurfaces) or can in turn be periodically ordered to construct periodic metasurfaces, where the particles that form the repeating unit cell are call meta-atoms. In this context, the term metasurface is coined for arrays with sub-wavelength lattice constant  where only the non-diffractive orders can propagate to the far field, whereas metagrating is used to indicate that diffraction can be relevant. 

The properties of a system can be fully described in terms of their resonant modes, where a resonant mode is a self-consistent solution of the electromagnetic field in the absence of external sources. For infinitely extended systems, as periodic planar arrays, the nature of the resonant modes can be divided in two kinds: leaky and confined modes. Typically, the resonant frequencies are complex  inside the continuum of radiation (for the in-plane component of the wavevector smaller than the free propagating wavevector), with an imaginary part proportional to the timescale at which the energy is leaked out to the far field. 
However, it is also possible to find confined modes, characterized by real frequencies, inside the continuum of radiation. Notably, despite there being available radiation channels in which the energy can decay, these modes cannot couple into them, receiving the name of bound states in the continuum (BICs).  
Thus, BICs are resonant states with infinitely high Q-factors that cannot be excited by far field radiation~\cite{Marinica2008,Lee2012b,Hsu2013,Hsu2016,Koshelev2018,Minkov2018,Doeleman2018,Koshelev2019a,Abujetas2019d,Abujetas2019c,Murai2020b,Chen2020a,Azzam2021}. Nonetheless, under perturbation in the parameter space around the BIC condition, it is possible (from the far field) to excite resonances (quasi-BICs) with arbitrary large Q-factors and huge enhancement of the electromagnetic field at the near field, holding promise (both BICs and quasi-BICs) of unprecedented planar devices in Nanophotonics. 
These interesting properties have been extensively investigated for diverse photonic applications such as enhanced sensing \cite{Yanik2011,Yesilkoy2019,Abujetas2019a,Tseng2020,Romano2020}, filtering \cite{Foley2014}, lasing \cite{Hsu2013,Kodigala2017,Ha2018,Khaidarov2019,Murai2020,Wu2020,Hwang2021}, electromagnetically-induced transparency \cite{Abujetas2021c}, chirality \cite{Gorkunov2021,Kim2021},  and non-linear conversion \cite{Carletti2018,Anthur2020,Bernhardt2020,Kang2021a}.

Depending on the mechanism that prevents the coupling of BIC with the continuum of radiation, they are classified as symmetry-protected or accidental BICs~\cite{Hsu2016,Azzam2021,Sadrieva2019}. Symmetry-protected BICs have been largely explored and they can only be found at the $\Gamma$ point, where there is a mismatch between the symmetry of the field mode and the available radiation channel imposed by the in-phase oscillation of the particles in the array. By contrast, accidental BICs emerge when different modes are coupled and they interfere destructively at a specific angle, typically arising in high-order resonant bands in photonic crystal slabs \cite{Hsu2013,Romano2020,Kodigala2017,Hwang2021,Sadrieva2019,Kang2021,Han2021}. Despite both photonic crystals and metasurfaces can be considered analogous in the non-diffracting regime as quasi-planar periodic arrangements, there are subtle theoretical differences in the manner that connected domains (slab with holes in photonic crystals) polarize light, as compared with isolated domains (meta-atoms in the case of metasurfaces) that allow for resonant electromagnetic confinement inside, with non-trivial consequences in the resulting phenomenology; and Babinet-like principles are not at all evident as long as holes/meta-atoms are thick and penetrable \cite{Ortiz2021}. In the case of metasurfaces, since particles in the array suffer from depolarization effects from the own lattice, the conditions for the formation of accidental BICs are not evident, and there are no guidelines to design them in the wavevector space; in fact, it has been assumed that multipolar contributions are needed to yield accidental BICs in  metasurfaces \cite{Sadrieva2019}, condition that we will demonstrate not strictly necessary, as in the case of one-dimensional grating  \cite{Doeleman2018}, with periodicity only along one in-plane direction, invariant along the other.

In this work, the emergence of accidental BICs in a rectangular metasurface of dielectric particles is investigated on the basis of a coupled electric and magnetic dipole (CEMD) formulation \cite{Abujetas2020a,Abujetas2021}. First, generic conditions are established, demonstrating that, for arrays of axially symmetric particles, accidental BIC can only arise along the symmetry lines $\Gamma X$ (and $\Gamma Y$ for rectangular arrays). Analytic expressions are then derived to determine the condition for accidental BICs  along these symmetry lines, showing that the individual polarizability of the particles must be related to the lattice depolarization Green function. These conditions only depend on the geometry of the lattice and meta-atom polarizibilities, enabling us to establish a general guideline for the formation of accidental BICs. Later, the accidental BIC condition for rectangular arrays of all-dielectric spheres is studied, giving examples of systems supporting TE and TM accidental BICs. Interestingly, it is shown that rectangular arrays can support more than one accidental BIC for the same metasurface at a given polarization. Finally, a feasible example with a square array of semiconductor nanodisks is presented, where the aspect ratio of the disk offers us a new degree of freedom for engineering the polarizability of the meta-atom. In addition, rectangular metasurfaces also allow  to tailor the emergence of, not only single, but also double (for both linear polarizations) accidental BICs at the same position in $\omega,k$-space (any off-normal angle of incidence along $\Gamma X$ or $\Gamma Y$ for a given frequency). This in turn confirmed through full numerical simulations, in good agreement with our quasi-analytical CEMD calculations. Although all the phenomenology is analyzed for rectangular arrays and one meta-atom per unit cell, the principles used in this study can be extended to metasurfaces with any geometry as long as the CEMD approach remains valid.

\section{CEMD formulation}

First, let us consider an infinite rectangular array of identical particles, labeled as $(n,m)$ and placed at
\begin{equation}
\r_{nm} = x_n \hat{x} + y_n \hat{y} = na \hat{x} + mb \hat{y},
\end{equation}
where $a$ and $b$ are the lattice constants along the $x$ and $y$ axis, respectively. 
For the homogeneous problem in the absence of external illumination, each particle in the array is excited by the waves emitted from the rest of the array.
The self-consistent incident field on the $(n,m) = (0,0)$ particle ($\r_{00} =  \bm 0$), $\bm{\Psi}_{\text{inc}}(\bm 0)$, is then given by the solution of 
\be
\bm{\Psi}_{\text{inc}}(\bm 0) 
=
\sum_{nm}{'}
k^2 \GG(-\r_{nm}) \alphagg
\bm{\Psi}_{\text{inc}}(\r_{nm}),
\ee
where $\sum_{nm}{'}$ means that the sum runs for all indices except for $(n, m) = (0, 0)$. $\GG(\r)$ and $\alphagg$ are matrices representing the dyadic Green function and the dipolar polarizability of the particles, respectively, and their representations depend on the  basis chosen to describe the electromagnetic fields, $\bm{\Psi}(\r)$. For the fields, the time dependence $\exp\left(-i\omega t\right)$ is assumed;  $\omega$ is the angular frequency, related to the modulus of the wavevector through $k = \omega/c$, $c$ being the speed of light. The dyadic Green function is obtained from the scalar Green function, $g\left(\r\right)$, by applying a linear differential operator, $\mathbf{\mathcal{L}}$, that also depends on the chosen basis~\cite{Abujetas2020a}

For periodic arrays the Bloch's theorem holds, $\bm{\Psi}_{\text{inc}}(\r_{nm})  = \bm{\Psi}_{\text{inc}}(\bm 0) \exp(\ii k_x n a)\exp(\ii k_y m b) = \bm{\Psi}_{\text{inc}}(\bm 0) e^{\ii\phi _{nm}}$, where $k_x$ and $k_y$ are the in-plane components of the wavevectors. Thus, the self-consistent incident field can be written as
\begin{align}
\bm{\Psi}_{\text{inc}}(\bm 0)
=
k^2\left[\sum_{nm}{'} \GG(-\r_{nm}) e^{\ii\phi_{nm}}\right] \alphagg \bm{\Psi}_{\text{inc}}(\bm 0) \equiv k^2 \GGL_{b}\alphagg \bm{\Psi}_{\text{inc}}(\bm 0).
\label{Eq:int}
\end{align}
We have defined $\GGL_{b}$, the lattice \textit{depolarization} dyadic (or return Green function), as 
\be
\GGL_{b}(k,k_x,k_y) \equiv \sum_{nm}{'}
\GG(-\r_{nm},k,k_x,k_y)e^{\ii\phi_{nm}(k,k_x,k_y)},
\label{eq:Gb_2D_ini}
\ee
and we have explicitly shown the dependence on both the modulus wavevector and the in-plane components of the wavevector, $\mathbf{k_{||}} = (k_x, k_y)$. $\GGL_{b}$ tells us about the coupling strength between particles, and is crucial to determine all the lattice properties and the nature of the modes supported by the metasurface.  Finally, we rearrange Equation~(\ref{Eq:int}) as follows:
\begin{equation}
\left[\II - k^2 \GGL_{b}\alphagg \right]
\bm{\Psi}_{inc}(\bm 0) = \bm 0,
\label{eq:cemd}
\end{equation}
where $\II$ is the unit dyadic.

In order to find the resonant states of the metasurfaces we need to find a solution to the homogeneous linear system of equations Equation~\eqref{eq:cemd}, appearing  only when the determinant is equal to zero:
\be
\left| \II - k^2\GGL_b\alphagg \right| = \left| \dfrac{1}{k^2\alphagg} - \GGL_b \right| = 0.
\label{eq:BS}
\ee
The complex frequencies at which Equation~\eqref{eq:BS} is satisfied are the eigenfrequencies, denoted by $\nu = \nu' - i\nu''$, where $\nu'$ and $\nu''$ are real numbers that denote the real and imaginary parts, respectively. In the latter eigenmode equation, it is more convenient 
to use the second expression rather than the first one because the imaginary part is well defined. Since $\GGL_{b}$ can be expressed in the reciprocal space~\cite{Abujetas2020a}, Equation~\eqref{eq:BS} can  thus be employed to determine the dispersion relation of resonant modes in metasurfaces. 
 For the sake of simplicity, we will assume in what follows that the array is embedded in a uniform medium (vacuum, indeed), so that $\GG$ in Equation (\ref{Eq:int}) is the free space Green dyadic. This is equivalent in an experimental configuration with a metasurface on a substrate to adding an index-matching layer \cite{Murai2020b}. Nonetheless, the impact of a substrate could be straightforwardly incorporated in our CEMD formulation by including the Green dyadic for two semi-infinite media.

 \section{Generic condition for accidental BICs }

In the more generic case (with any restriction for the values of $\bf{k_{||}}$), the determinant can be only factorized into two terms (for diagonal polarizabilities), one for each polarization:
\be
\left| \dfrac{1}{k^2\alphagg} - \GGL_b \right| = \left|\eta^{(TE)}_{xyz}\right| \left| \eta^{(TM)}_{xyz} \right|,
\label{eq:eingen2}
\ee
with
\be
\eta_{xyz}^{(TE)} &=& \left(\dfrac{1}{k^2 \alpha_{x}^{(e)}} - G_{bxx}  \right)\left(\dfrac{1}{k^2 \alpha_{y}^{(e)}} - G_{byy}  \right)\left(\dfrac{1}{k^2 \alpha_{z}^{(m)}} - G_{bzz}  \right) \nonumber \\
&-& \dfrac{1}{k^2 \alpha_{x}^{(e)}}G_{byz}^2 - \dfrac{1}{k^2 \alpha_{y}^{(e)}}G_{bzx}^2 - \dfrac{1}{k^2 \alpha_{z}^{(m)}}G_{bxy}^2 + 2G_{bxy}G_{byz}G_{bzx},
\nonumber \\
\eta_{xyz}^{(TM)} &=& \left(\dfrac{1}{k^2 \alpha_{x}^{(m)}} - G_{bxx}  \right)\left(\dfrac{1}{k^2 \alpha_{y}^{(m)}} - G_{byy}  \right)\left(\dfrac{1}{k^2 \alpha_{z}^{(e)}} - G_{bzz}  \right) \nonumber \\
&-& \dfrac{1}{k^2 \alpha_{x}^{(m)}}G_{byz}^2 - \dfrac{1}{k^2 \alpha_{y}^{(m)}}G_{bzx}^2 - \dfrac{1}{k^2 \alpha_{z}^{(e)}}G_{bxy}^2 + 2G_{bxy}G_{byz}G_{bzx}.
\label{eq:aBg}
\ee
In these expressions, $G_{bij}$ are the matrix elements of $\GGL_{b}$, $\alpha_i^{(e),(m)}$ are the diagonal matrix elements of the electric $(e)$ and magnetic $(m)$ polarizability, and $TM$, $TE$ stand for transverse electric and magnetic modes, respectively. To simplify, the upper index $^{(EM)}$ used in ref.~\cite{Abujetas2020a} (indicating that they connect electric and magnetic dipoles), is omitted in $G_{byz}$ and $G_{bzx}$, understood for physical reasons.
Note that $G_{bxy}$ couples either electric only or magnetic only dipoles, while $G_{byz}$ and $G_{bzx}$ connect electric with magnetic dipoles. Also, for these expressions, the sign of $G_{byz}$ and $G_{bzx}$ is taken in such a way that:
\be
\Imag\left[G_{byz}\right] = -\dfrac{1}{2abk_z} \dfrac{k_x}{k}, \quad 
\Imag\left[G_{bzx}\right]  = -\dfrac{1}{2abk_z} \dfrac{k_y}{k},
\ee
since their signs must be consistent with their definitions.

The resonant surface modes arising from the zeroes of $\eta^{(TE)}_{xyz}$ (respectively, $\eta^{(TM)}_{xyz}$) represent hybrid modes with different polarization, where the electric (respectively, magnetic) in-plane dipoles are coupled among them (by $G_{bxy}$) and with the magnetic (respectively, electric) dipoles along the $z$ axis (by $G_{byz}$ and $G_{bzx}$). The interference between the different dipolar modes leads to interesting phenomenology, with the open question yet on whether accidental BICs can be supported: recall that the imaginary part of the above mentioned resonant modes must vanish ($\nu''=0$) to be considered BICs.

To this end, let us analyze Equations~(\ref{eq:aBg}) in a general manner. First, note that it is possible to solve for the zeroes of each of  Equations~(\ref{eq:aBg})  as a system of two equations for the real and imaginary parts (separately) with three unknowns: the real parts of the inverse of the polarizabilities along the three axis. Then the accidental BIC conditions can be solved for either mode as a function of a parameter (the real part of the inverse of one of the polarizabilities), leading to plausible solutions (zeroes) of Equations~(\ref{eq:aBg})  along any direction in $k$-space by tuning the remaining polarizibilities. This is true for non axially symmetric meta-atoms (as an ellipsoidal disk or a regular disk with its axis in the $xy$ plane) with three different polarizibilities along  the main axes.

Nonetheless, if the particles possess axial symmetry around the $z$ axis (as in a disk), in-plane polarizibilities are equal, $\alpha_x = \alpha_y$, and less degrees of freedom are available. In fact, it can be shown that, if both $k_x \not = 0$ and $k_y \not = 0$,  no real values for the inverse of the polarizability can be found for which Equation~(\ref{eq:eingen2}) is fulfilled. Therefore, it is not possible to design accidental BIC away from the symmetry lines $\Gamma X$ and $\Gamma Y$ (for rectangular arrays) for axially symmetric arrays of particles (in the dipolar regime). Moreover, no accidental BIC can be formed either along the symmetry lines $\Gamma M$ and $\Gamma S$. 

By contrast, $G_{bxy}$ and $G_{bzx}$ (or $G_{byz}$) become zero along the symmetry line $\Gamma X$ (respectively, $\Gamma Y$), and only the interference between the electric dipole along the $y$ axis (respectively, $x$ axis) and the magnetic dipole along the $z$ axis is relevant (and vice versa exchanging electric/magnetic). 
Therefore, since the degrees of freedom remain the same, polarizibilities satisfying Equation~(\ref{eq:eingen2}) can be found, so that accidental BICs along the symmetry line $\Gamma X$ (or $\Gamma Y$) can be supported by metasurfaces with axially symmetric  meta-atoms, as we reveal in the following sections. 

\begin{figure}
\includegraphics[width=1\columnwidth]{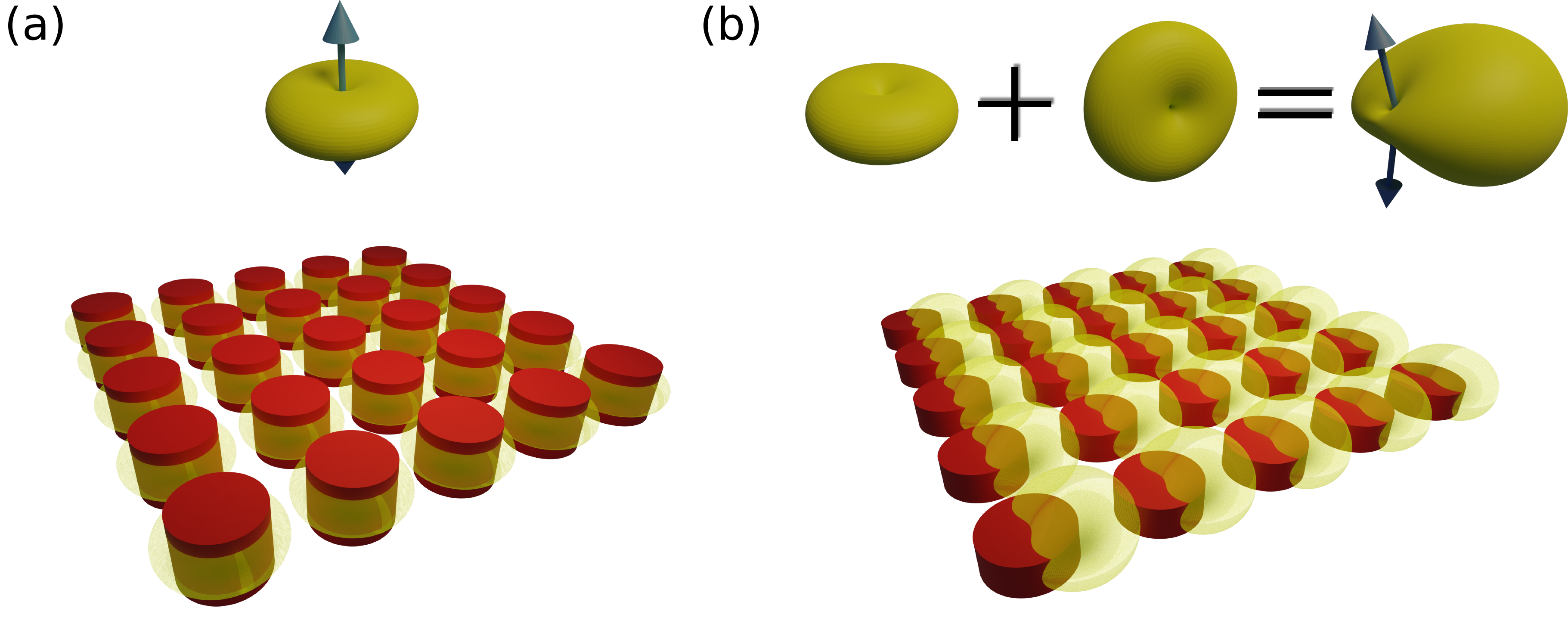}
\caption{Sketch of BICs supported by metasurfaces with one (dipolar) meta-atom per unit cell. (a) Symmetry-protected BIC formed by the in-phase oscillation of dipoles along the $z$ axis. (b) The destructive interference between in-plane electric (magnetic) dipoles and out-plane magnetic (electric) dipoles allows for the formation of accidental BICs. The arrows indicate the directions along which the emission pattern vanishes.}
\label{fig_1}
\end{figure}

\section{Accidental BICs along $\Gamma X$}

If we thus consider modes propagating along the $x$ axis ($k_x \not = 0$ and $k_y = 0$) and diagonal polarizabilities, 
Equation~\eqref{eq:BS} can be factorized in four terms~\cite{Abujetas2020a}:
\be
\left| \dfrac{1}{k^2\alphagg} - \GGL_b \right| = \left|\eta^{(TE)}_{x}\right| \times \left| \eta^{(TM)}_{x}\right| \times \left|\eta^{(TE)}_{yz}\right| \times  \left| \eta^{(TM)}_{yz} \right|,
\label{eq:eingen}
\ee
with 
\be
\eta^{(TE)}_{x} = \dfrac{1}{k^2 \alpha_{x}^{(e)}} - G_{bxx} , \quad  \eta^{(TE)}_{yz} = \left(\dfrac{1}{k^2 \alpha_{y}^{(e)}} - G_{byy}  \right)\left(\dfrac{1}{k^2 \alpha_{z}^{(m)}} - G_{bzz}  \right)  - G_{byz}^2, \nonumber \\
\eta^{(TM)}_{x} = \dfrac{1}{k^2 \alpha_{x}^{(m)}} - G_{bxx}, \quad  \eta^{(TM)}_{yz} = \left(\dfrac{1}{k^2 \alpha_{y}^{(m)}} - G_{byy}  \right)\left(\dfrac{1}{k^2 \alpha_{z}^{(e)}} - G_{bzz}  \right)  - G_{byz}^2.
\label{eq:eingen_s}
\ee
For waves propagating along the $y$ axis ($k_x = 0$ and $k_y \not = 0$), the phenomenology is the same by replacing the index $y$ by $x$ in Equation~(\ref{eq:eingen_s}).
Among the different resonant surface modes represented by each term in Eq.~\eqref{eq:eingen}, we exclude  $\eta^{(TE)}_{x}$ and $\eta^{(TM)}_{x}$ since they have been shown to yield no BIC whatsoever (all their poles inside the continuum of radiation have non-negligible widths,  $\nu''\not =0$). 

We thus  focus on the hybrid modes, $\eta^{(TE)}_{yz}$ and $\eta^{(TM)}_{yz}$, emerging from electric (magnetic) dipoles along the $y$ axis coupled to magnetic (electric) dipoles along the $z$ axis. At the $\Gamma$ point, $k_x=k_y=0$, well known symmetry-protected BICs arise~\cite{Abujetas2020a}. For the sake of completeness, we explicitly show the conditions, not shown in ref.~\cite{Abujetas2020a}, yielding symmetry-protected BICs given by the in-phase oscillation of lossless dipoles along the $z$ axis:
\be
\begin{array}{cc}
\mathrm{TE \ symmetry-protected \ BIC} \quad & \quad \mathrm{ TM\ symmetry-protected \ BIC}  \\
\Real\left[\dfrac{1}{k^2 \alpha_{z}^{(m)}}\right]_{sBIC} = \Real\left[G_{bzz}\right], \quad & \quad \Real\left[\dfrac{1}{k^2 \alpha_{z}^{(e)}}\right]_{sBIC} = \Real\left[G_{bzz}\right].
\end{array}
\label{sym_BIC}
\ee 
Therefore, for a metasurface to support symmetry-protected BICs, the real part of the inverse of the polarizability of the meta-atoms must be consistent with the lattice properties.

Searching for  accidental BICs at $k_x\not =0$ (with $k_y = 0$), also for lossless meta-atoms, the condition $\eta_{yz}=0$ yields a system of two equations (real and imaginary parts) with two unknowns: the real parts of the inverse of the polarizabilities along the $y$ and $z$ axes. After a straightforward derivation, the accidental BIC conditions are obtained:
\be
\begin{array}{cc}
\mathrm{TE \ accidental \ BIC} 
& \mathrm{TM \ accidental \ BIC}  
\\
\Real\left[\dfrac{1}{k^2 \alpha_{y}^{(e)}} \right]_{aBIC} = \Real\left[ G_{byy} + \dfrac{k}{k_x}G_{byz} \right], & \Real\left[\dfrac{1}{k^2 \alpha_{y}^{(m)}} \right]_{aBIC} = \Real\left[ G_{byy} + \dfrac{k}{k_x}G_{byz} \right],\\
\Real\left[\dfrac{1}{k^2 \alpha_{z}^{(m)}} \right]_{aBIC} = \Real\left[ G_{bzz} + \dfrac{k_x}{k}G_{byz} \right], & \Real\left[\dfrac{1}{k^2 \alpha_{z}^{(e)}} \right]_{aBIC} = \Real\left[ G_{bzz} + \dfrac{k_x}{k}G_{byz} \right];
\end{array}
\label{acc_BIC}
\ee
note that two conditions, for $1/\alpha_{y}$ and $1/\alpha_{z}$ separately,  must be satisfied simultaneously. 

Equations~(\ref{sym_BIC}) and (\ref{acc_BIC}) give us an interesting perspective about the differences between symmetry-protected and accidental BICs. Symmetry-protected BICs only need to fulfil one condition at a specific frequency for the polarizability (with in-plane modulus wavevector $k_{||} = 0$), whereas accidental BICs need to fulfil two conditions at the same frequency and longitudinal component of the wavevector. Then, accidental BICs obey more restrictive conditions that are not always reachable, depending on the precise cancellation of radiation from different dipolar modes. These characteristics are illustrated in Figure~\ref{fig_1}: a symmetry-protected BIC may simply arise from vertical dipolar resonances, whereas an accidental BIC requires the  interference between different in-plane and out-of-plane dipolar modes to cancel radiation to the far-field at specific off-normal angles. Quite importantly, Equation~(\ref{acc_BIC}) is a guideline to engineer accidental BICs for a wide variety of all-dielectric (and plasmonic) photonic metasurfaces with a common (rectangular) lattice symmetry.

\section{Accidental BICs on all-dielectric nanosphere metasurfaces}

\begin{figure}
\includegraphics[width=1\columnwidth]{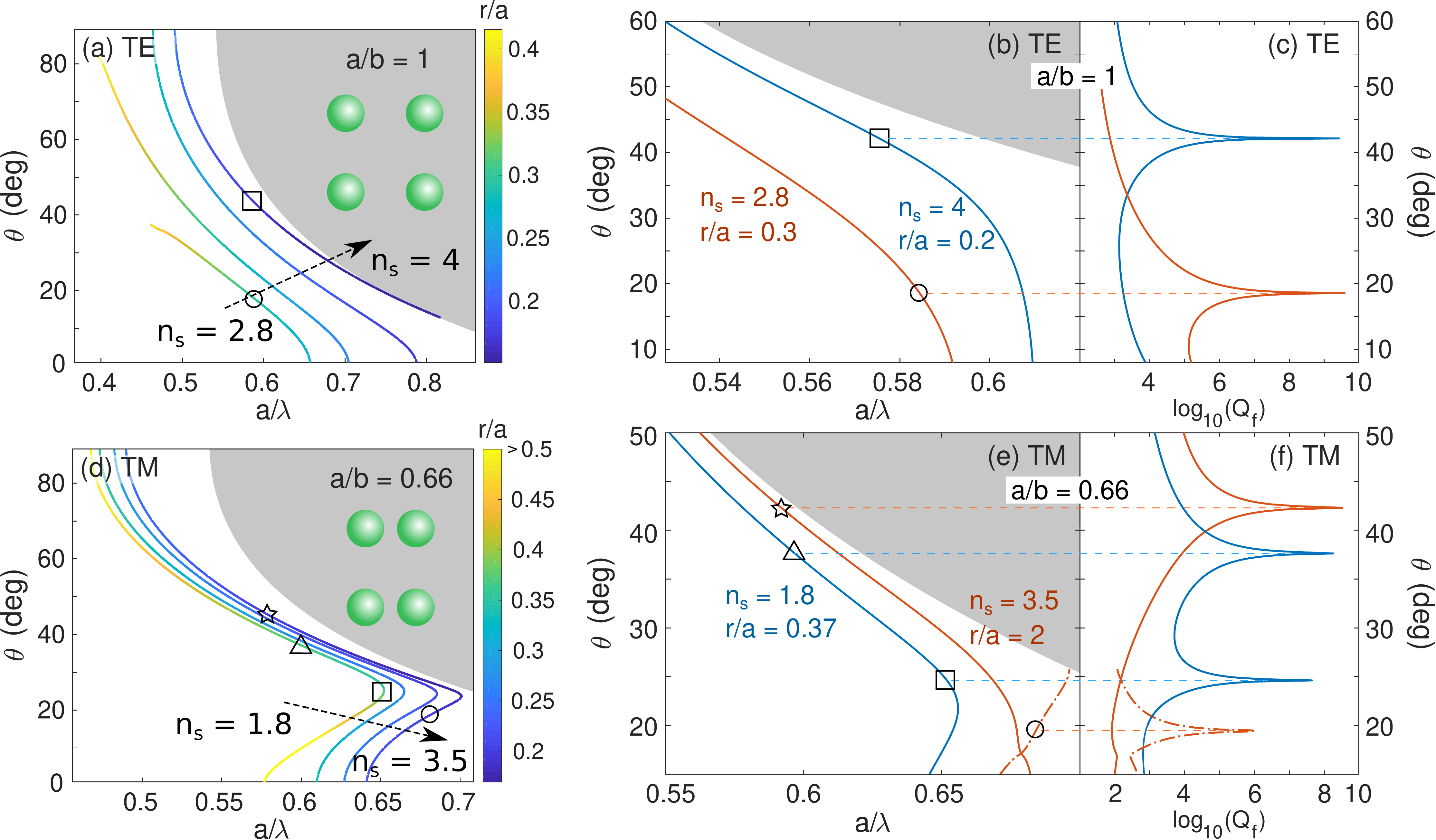}
\caption{
Accidental BIC condition in parameter space $(a/\lambda,\theta)$ for: (a) a square array ($a = b$) in TE polarization at $n_s =$ 2.8, 3, 3.5 and 4; (d) a rectangular array ($a/b =$ 0.66) in TM polarization at $n_s =$ 1.8, 2.5, 3 and 3.5. Color lines indicate fixed $r/a$ ratios and the shadowed sections correspond to diffracting spectral regions. Dispersion relation of modes that evolve to  accidental BICs, marked with geometric symbols in (a, d), for (b) TE polarization and (e) TM polarization. (c, f) Quality factors of the resonant modes defined as $Q_f = \nu'/\nu''$, revealing the divergence associated to accidental BIC emergence.
}
\label{fig_2}
\end{figure}

To shed light on the phenomenology resulting from the previous theoretical formulation, we represent in Figure~\ref{fig_2} the accidental BIC position as a function both of the normalized frequency ($a/\lambda$) and of the angle of incidence ($\theta$) for rectangular arrays (lattice parameters $a,b$) of dielectric spheres with different values of their refractive index. The polarizability of the spheres is calculated from Mie theory, taking into account only the first two (dipolar) terms of the harmonic expansion. The color of each line is correlated with the normalized sphere radius $r/a$ that the metasurface must fulfil in order to support an accidental BIC, where $r/a = 0.5$ represents a metasurface of touching spheres. The results for TE accidental BICs in square arrays ($a,b$) are analyzed in Figure~\ref{fig_2}a. It can be seen that the general tendency is that a higher $r/a$ ratio is needed to support accidental BICs as the refractive index of the particles becomes smaller and as the longitudinal component of the wavevector increases. Also, note that the curves corresponding to $n_s = 2.8$ and $n_s = 3$ end before achieving the condition $k = k_x$ (being $k_x$ related to $\theta$ through the expression $k_x = k\sin\theta$). 
Indeed, it can be found that no accidental BIC can be formed for approximately $n_s < 2.5$. On the other hand, as the value of $n_s$ increases, accidental BICs  become only accessible at higher values of $k_x$, reaching the diffraction limit, as illustrated in Figure~\ref{fig_2}a for $n_s = 4$.

By contrast, we have found no accidental BICs in TM polarization for square arrays of all-dielectric spheres. For dielectric spheres, the dipolar mode $\alpha_y^{(m)}$ resonates at  frequencies lower than that of  $\alpha_z^{(e)}$, precluding its formation according to Equations~\eqref{acc_BIC}. Alternatively, we study TM accidental BICs in rectangular arrays that indeed support them, as revealed in Figure~\ref{fig_2}d for $a/b = 0.66$.
As in the case of TE polarization, larger spheres are needed for lower refractive indices.
Interestingly, these rectangular metasurfaces support accidental BICs for dielectric spheres with relative small values of $n_s$; namely, high-refractive indices are not strictly necessary, which widens the range of dielectric materials that could be exploited in this respect. For example, accidental BICs can be engineered  around $\theta = 20^{\circ} - 50^{\circ}$ for spheres with $n_s = 1.8$ and meaningful radii. For higher values of $n_s$, accidental BICs arise at higher frequencies, being possible to engineer them for $n_s \approx 5$, but they are not accessible for all values of $\theta$. Also, it is remarkable that two accidental BICs can be found in parameter space for the same $n_s$ and $r/a$ parameters (i.e., for the same metasurface). For example, there are accidental BICs at $(a/\lambda, \theta) \approx (0.68, 19)$ and $(a/\lambda, \theta) \approx (0.59, 42)$  for $n_s = 3.5$ and $r/a = 0.2$. 

By way of example, we show in Figure~\ref{fig_2} the dispersion relations and  associated quality factors for modes approaching the accidental BIC condition in four specific metasurfaces.  These magnitudes are calculated by solving Equation~(\ref{eq:BS}) and using the formalism described in Ref. \cite{Abujetas2020a}. The dispersion relations $\nu'(k_x)$ are shown in the parameter space given by  angle $\theta$ (recall that $k_x=k\sin\theta$) and normalized frequency $a/\lambda$ (with $\lambda = c/\nu'$); the quality factor is defined as $Q_f = \nu'/\nu''$ and calculated along the mode dispersion. TE modes in square arrays are shown in Figure~\ref{fig_2}b,c for two different metasurfaces: high-refractive index particles $n_s = 4$ and $r/a = 0.2$ (blue curves); moderate high-refractive index particles $n_s = 2.8$ and $r/a = 0.3$ (red curves). As expected, the quality factor of the modes diverges at the predicted position in parameter space of the accidental BICs, around $\theta = 43^{\circ}$ and $\theta = 18^{\circ}$ for the two cases shown in Figure~\ref{fig_2}b,c. The dispersion relations of TM modes in rectangular arrays ($a/b = 0.66$) are also represented in Figure~\ref{fig_2}e,f. The blue curves correspond to spheres with low index of refraction, $n_s = 1.8$ and $r/a = 0.37$, whereas red curves are for high refraction index particles, $n_s = 3.5$ and $r/a = 0.2$. Interestingly,  both metasurfaces support two accidental BICs for the same polarization, and their relative distance in parameter space can be tailored by the $r/a$ ratio. Incidentally, if  $r/a$ is decreased, a threshold value is found at which both accidental BICs collapse into a single one at the same angle and frequency, below which no accidental BIC is found (although a narrow resonance is still present). 

\begin{figure}
\includegraphics[width=1\columnwidth]{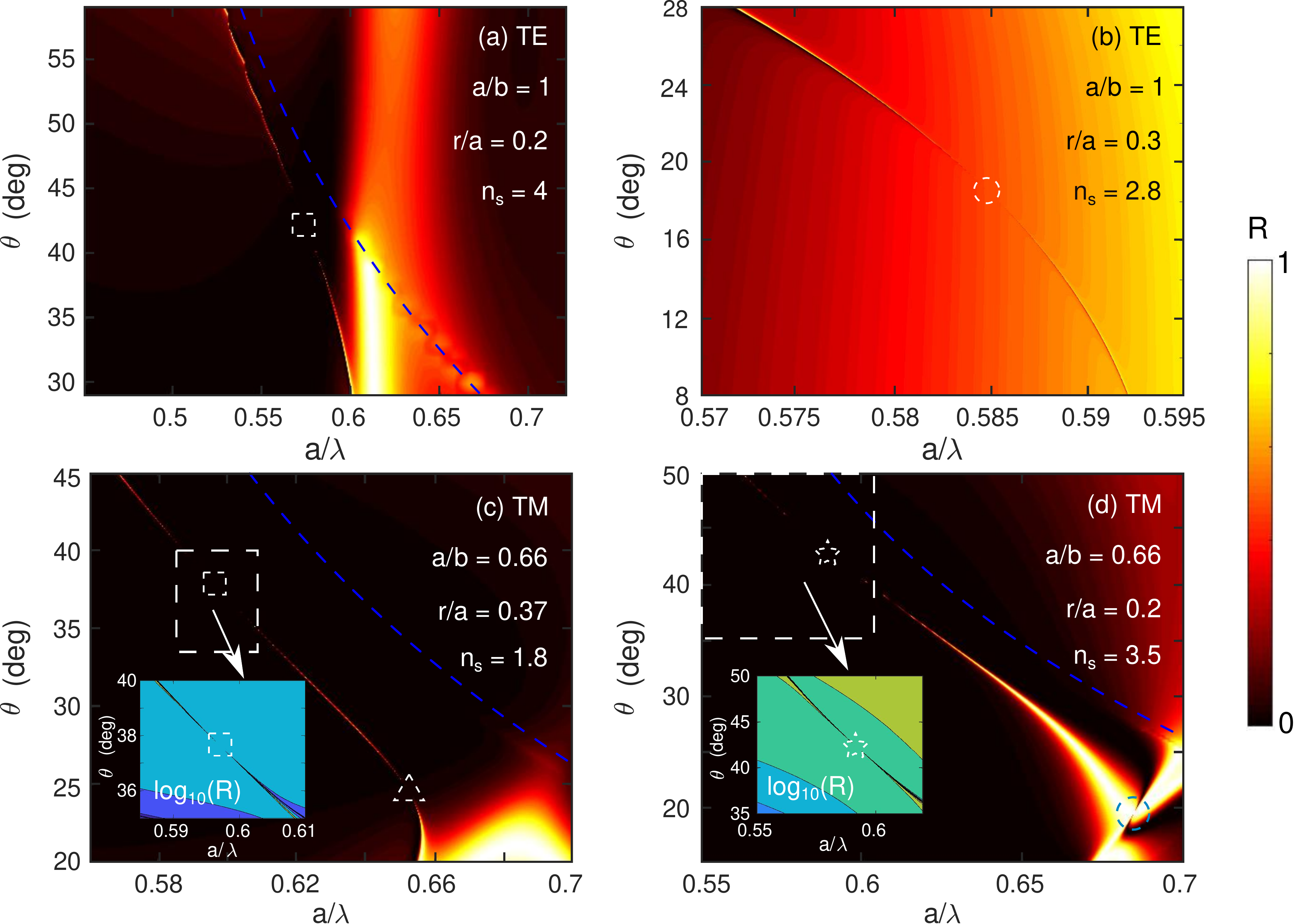}
\caption{Color maps of the reflectance as a function of normalized frequency and angle of incidence for four particular cases, marked with  geometric symbols in Figure~\protect{\ref{fig_2}}b,e, with the following parameters: TE polarization for a square array with (a) $n_s = 4$ and $r/a = 0.2$; (b) $n_s = 2.8$ and $r/a = 0.3$; TM polarization for a rectangular array ($a/b = 0.66$) with (c) $n_s = 1.8$ and $r/a = 0.37$; (d) $n_s = 3.5$ and $r/a = 0.2$.  
}
\label{fig_3}
\end{figure}

For the sake of illustration, the reflectance is shown in Figure~\ref{fig_3} for different configurations supporting accidental BICs, marked as black symbols in Figure~\ref{fig_2}a,e. The reflectance maps have been calculated using the CEMD formulation \cite{Abujetas2020a}. For TE polarization, two square arrays have been considered in Figure~\ref{fig_3}a,b. First, an array of spheres with high refractive index $n_s = 4$ and radius $r/a = 0.2$ is examined in Figure~\ref{fig_3}a. The expected behavior in parameter space (formally equivalent to the $\omega,k$) is observed, associated with a narrow (accessible) quasi-BIC band with high Q-factors turning into an (inaccessible) BIC, with a vanishing reflectance corresponding to the BIC at the predicted angle around $\theta = 43^{\circ}$. The other example is shown in Figure~\ref{fig_3}b for a lower (but moderately high) refractive index of  $n_s = 2.8$ at $r/a = 0.3$, for which the accidental BIC appears around $\theta = 18^{\circ}$ over a broad high reflectance background. 

For TM polarization, we have considered the two examples as marked as black symbols in Figure~\ref{fig_2}e. First, the interesting case mentioned above for a low index of refraction ($n_s = 1.8$) and  $r/a = 0.37$: as shown in Figure~\ref{fig_3}c, two accidental BICs are found in a rectangular array ($a/b = 0.66$) around $\theta \approx 25^{\circ}$ and $38^{\circ}$ . Note that there are very narrow high-reflectance bands between both BIC conditions, barely visible. Indeed, the accidental BIC occurring at large angles is zoomed in in the inset of Figure~\ref{fig_3}c.

Finally, Figure~\ref{fig_3}d presents the reflectance for the rectangular array of high refractive index spheres with $n_s = 3.5$ and $r/a = 0.2$. This metasurface also supports two accidental BICs with an interesting phenomenology. The accidental BIC around $\theta \approx 42^{\circ}$ shows the typical vanishing feature, i.e., there is no background. Conversely, the BIC around $\theta \approx 19^{\circ}$ occurs over a broad reflectance background with a complex pattern stemming from the combination of a broad resonance; then the quasi-BIC to BIC transition appears as a narrow dip within the broad band, stemming from the interference that fades away within the high reflectance  background, as a BIC-induced transparency band \cite{Abujetas2021c}. 

Although all preceding cases have been discussed in normalized frequencies (which could be easily extrapolated to any spectral regime), we would like to emphasize that all of them can be associated to realistic nanosphere metasurfaces supporting accidental BICs in the optical domain,  e.g., with lattice parameters of the order of $a=300$ nm, and dielectric spheres made of low (oxides) or high (semiconductors) refractive indices. To further stress the ability of Equations~\eqref{acc_BIC} to design all-dielectric metasurfaces with accidental BICs, we apply it next to a realistic scenario similar to that already used for symmetry-protected BICs \cite{Murai2020}.

\section{Accidental BICs on semiconductor nanodisk metasurfaces}

\begin{figure}
\includegraphics[width=1\columnwidth]{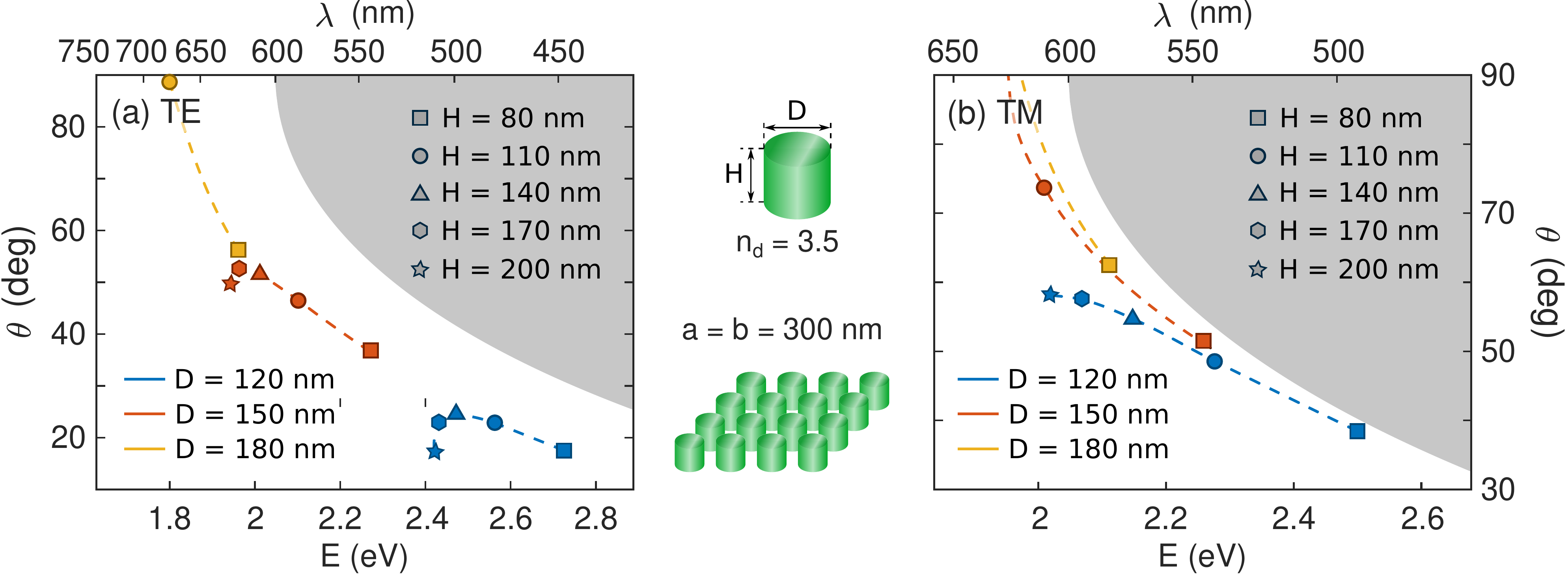}
\caption{Accidental BIC condition in parameter space for square arrays of semiconductor nanodisks with refractive index $n_d = 3.5$, as a function of the diameter, $D$, and height, $H$. The lattice constants of the array are $a = b =$ 300 nm: (a) TE polarization; (b) TM polarization. The dashed curves are calculated by interpolating the polarizabilities for different heights at a constant diameter, as a guide to the eye. The shadow sections correspond to diffracting spectral regions.}
\label{fig_4}
\end{figure}

In order to give a broader view of accidental BICs in all-dielectric metasurfaces, the phenomenology is also investigated  for a more practical case: a square array of dielectric nanocylinders of diameter, $D$, and height, $H$. Moreover, the additional degree of freedom, given by the aspect ratio of the particle, offers  a new parameter to engineer the properties of accidental BICs. Unlike for spherical particles, the relative spectral position of the resonances can be tuned with the aspect ratio of cylindrical particles. To this end, the polarizabilities of the disks are numerically calculated through SCUFF \cite{SCUFF1,SCUFF2}, rigorously accounting for the different components \cite{Bobylev2020}; a refractive index of $n = 3.5$ is assumed, as a typical value for dielectric semiconductors at optical and near-IR frequencies \cite{Murai2020}. Specifically, the polarizabilities of disks with diameters $D = 120, 150$, $180$, and $210$ nm, and heights $H = 80, 110, 140, 170,$ and $200$ nm (with all different combinations) are calculated.

The accidental BIC position for an array of semiconductor disks with lattice constant $a = b = 300$ nm (square array) is presented in Figure~\ref{fig_4}. Therein the position of the symbols marks the point in the parameter space at which the metasurface supports an accidental BIC for the different disks calculated numerically through Equations~\eqref{acc_BIC}. In addition, the dashed lines indicate the evolution of the spectral position as a function of the diameter, where the polarizability is calculated by interpolating those for different heights at constant diameter. As a general trend, the angular position of the accidental BIC  increases with the diameter in both polarizations, up to a point for larger disks that the conditions are no longer fulfilled. In addition, it should be emphasized that accidental BICs can be supported in square arrays of nanodisks for TM polarization, in contrast to the case of a square array of nanospheres. 

\begin{figure}
\includegraphics[width=1\columnwidth]{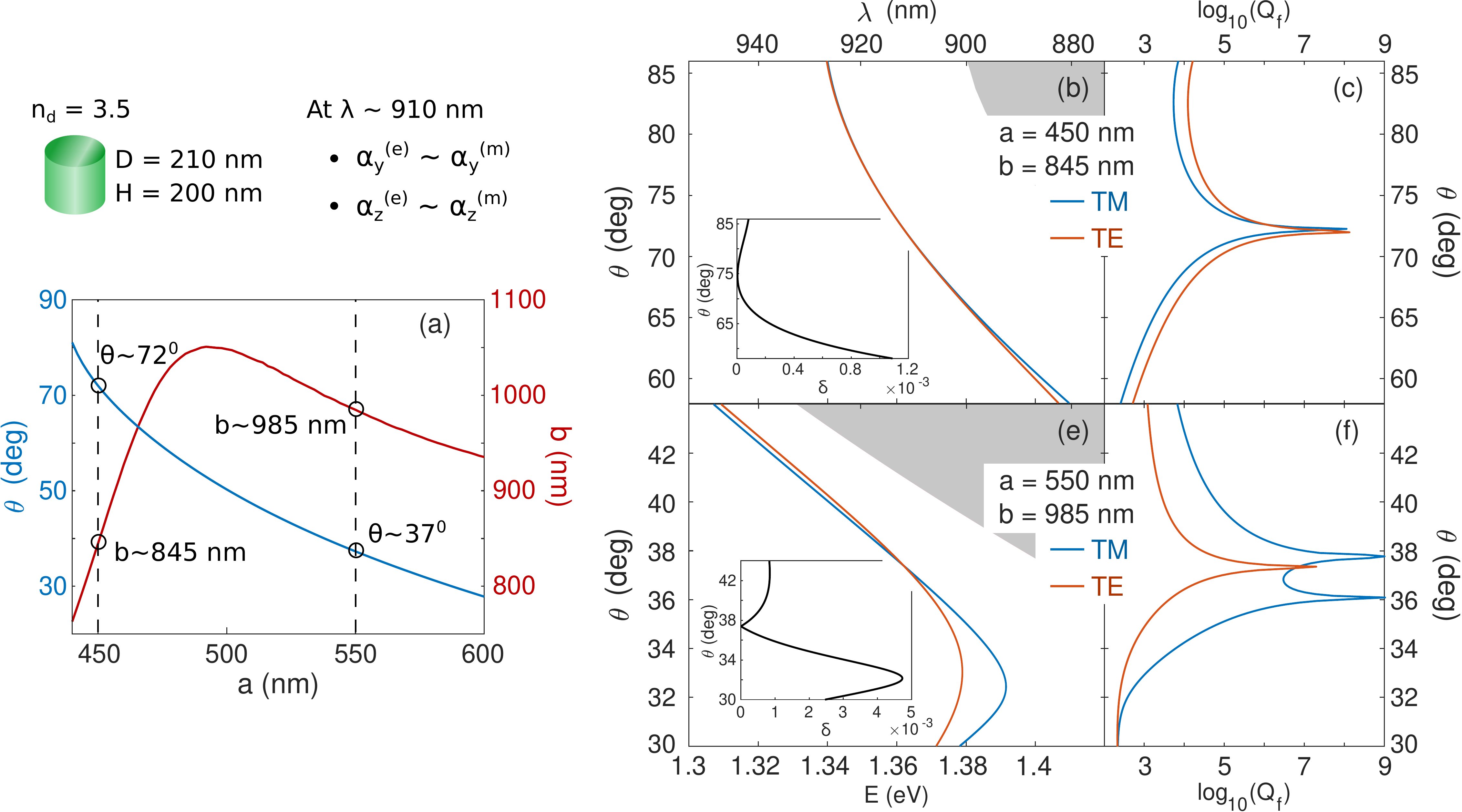}
\caption{(a) Condition for the emergence of double accidental BICs  for a rectangular array of semiconductor ($n_d=3.5$) nanodisks of diameter $D = 210$ nm and height $h = 200$ nm, as a function of $\theta$ and $b$ for different values of $a$. The dashed curves correspond to the cases at which the relation dispersion and the reflectance is calculated. Dispersion relation at both polarization for (b) $a = 450$ nm and $b = 845$ nm and (d) $a = 550$ nm and $b = 985$ nm. The insets show the detuned between the resonances, $\delta$. The associated quality factors are shown in (c, e).
}
\label{fig_5}
\end{figure}
The appearance of accidental BICs for TM polarization is possible due to the tunability of the relative spectral position of the dipolar resonances of the individual nanodisks with different aspect ratios. Moreover, exploiting such versatility, it is also possible to find configurations at which the metasurface supports accidental BICs for both polarizations at the same position in the parameter space. The requirements are: (i) equal electric and magnetic in-plane polarizabilities, (ii) equal electric and magnetic out-of-plane polarizabilities; and (iii) fulfilment of Equations~\eqref{acc_BIC}. We have checked that a cylinder of $D = 210$ nm and $H = 200$ nm closely meets (i) and (ii) around the wavelength $\lambda \sim 910$ nm, where the parameters of the lattice can be tuned to fulfil the third requirement. Therefore, based on the properties of this specific disk and fixing the operation wavelength at $\lambda = 910$ nm, the double accidental BIC condition is found as a function of the lattice parameter along the $y$ axis, $b$, and the in-plane wavevector along the $x$ axis, $k_x$ (where $k_x = k\sin\theta$), by changing the lattice parameter along the $x$ axis, $a$. The results are illustrated in Figure~\ref{fig_5}a. As the lattice parameter $a$ becomes smaller, the angle at which the BIC is found moves to higher angles of incidence, up to the point below $a \sim 420$ nm such that no accidental BIC emerges (not shown in the graph). Also, it can be seen that the lattice parameter $b$ must be always bigger than $a$.  

Now let us calculate the dispersion relation  of the resonant modes in two specific configurations, proceeding as in the case of sphere arrays shown in Figure~\ref{fig_2}b,e. Note that, for a proper calculation of the dispersion relation, the polarizability at complex frequencies is needed, trivial for the analytical polarizabilities of spheres, but not for those of disks. Nonetheless, since we are interested in the dispersion relation around the accidental BIC, the imaginary part of the mode frequency, $\nu$'', is vanishingly small. Therefore, the polarizability at complex frequencies can be safely approximated to the polarizability at the real part of the complex frequency. This approximation is very accurate for modes with $Q_f > 10^3$.

First, Figure~\ref{fig_5}b,c shows the dispersion relations and quality factors for both polatizations for a metasurface with lattice parameters $a = 450$ nm and $b = 845$ nm. Both dispersion relations nearly overlap in parameter space. For the shake of clarity, the inset shows the detuning parameter between resonances, $\delta$, defined as $\delta = |\nu'_{TM}-\nu'_{TE}|/(\nu'_{TM}+\nu'_{TE})$:  note that $\delta < 2\times 10^{-5}$ in the range $\theta = 71^{\circ} - 78^{\circ}$. Also, it is remarkable that the quality factor of both modes diverges around $\theta = 72^{\circ}$, as predicted. A similar phenomenology is observed in a rectangular array with lattice constant $a = 550$ nm and $b = 985$ nm. As illustrated in Figure~\ref{fig_5}d,e, both modes lie down again close to each other in parameter space, although in this configuration the detuning is higher in the vicinity of the minimum. Nonetheless, around $\theta = 37^{\circ}$, where the quality factor becomes very large, the detuning is near zero. Significantly, the quality factor of the TM mode exhibits two divergences, associated to two accidental BICs, a phenomenology that was also shown above for dielectric sphere arrays in Figure~\ref{fig_2}f.

\begin{figure}
\includegraphics[width=1\columnwidth]{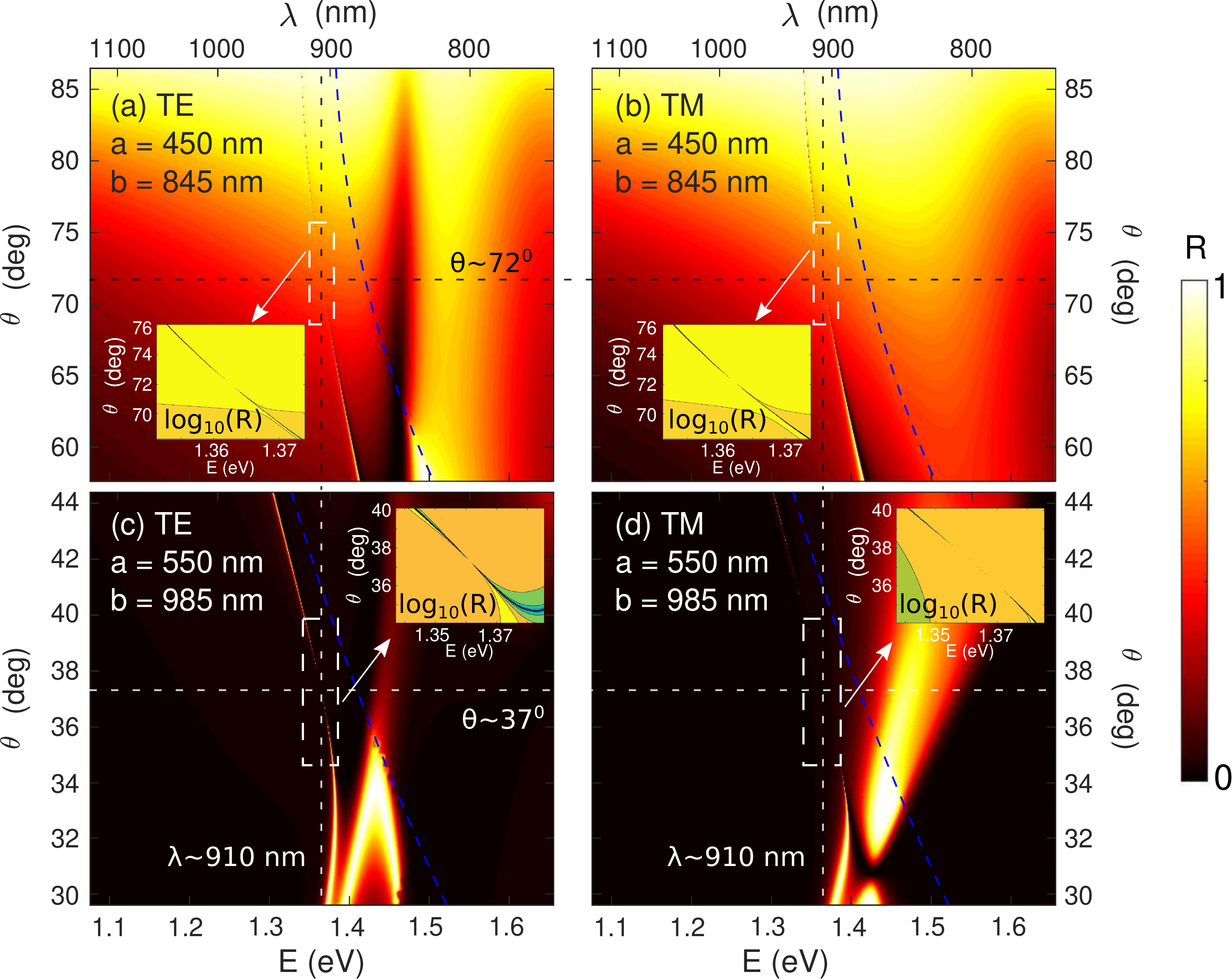}
\caption{Reflectance color maps for (a,c) TE and (b,d) TM polarization for rectangular arrays of Si nanodisks ($D = 210$ nm and $H = 200$ nm) with lattice parameters: (a,b) $a = 450$ nm and $b = 845$ nm, supporting a double accidental BIC at $\theta \sim 72^{\circ}$ and $\lambda \sim 910$ nm; (c,d)  $a = 550$ nm and $b = 985$ nm, supports a double accidental BIC at $\theta \sim 37^{\circ}$ and $\lambda \sim 910$ nm. Insets zoomed in the spectral region of the BIC conditions. Black or white horizontal/vertical dashed lines mark the BIC angular/spectral position. Blue dashed lines delimit the diffractive region. }
\label{fig_6}
\end{figure}

By choosing $a = 450$ nm, the specular reflectance is represented in Figures~\ref{fig_6}a,b at the double accidental BIC condition. For both polarizations, the accidental BIC is found at $\theta \sim 72^{\circ}$ (and at $\lambda \sim 910$ nm) when the lattice constant along the $y$ axis is $b = 845$ nm. Moreover, Figure~\ref{fig_6}c,d shows the reflectance at both polarizations for a rectangular array with lattice constants $a = 550$ nm and $b = 985$ nm. As predicted by Figure~\ref{fig_5}a, a double accidental BIC is supported at $\theta \sim 37^{\circ}$.

\begin{figure}
\includegraphics[width=1\columnwidth]{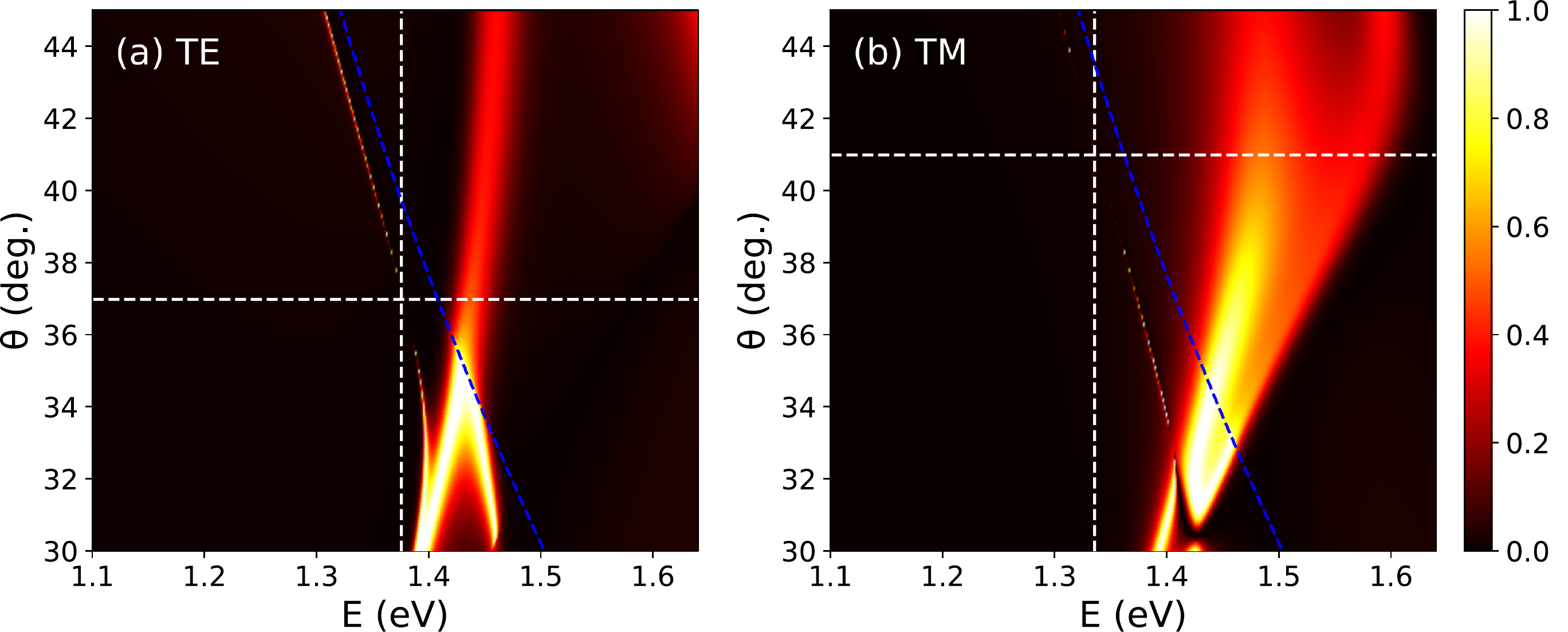}
\caption{  Numerically calculated (SCUFF) reflectance color maps  for (a) TE and (b) TM polarization, for the same case as that theoretically calculated  through our CEMD in  Figure~\protect{\ref{fig_6}c,d}: namely, a rectangular array of  Si nanodisks ($D = 210$ nm and $H = 200$ nm) with lattice parameters $a = 550$ nm and $b = 985$ nm that supports a double accidental BIC. White horizontal/vertical dashed lines mark the BIC angular/spectral position. Blue dashed lines delimit the diffractive region. }
\label{fig_7}
\end{figure}

Finally, we verify the accuracy of our theoretical model by carrying out numerical calculations through  SCUFF \cite{SCUFF1} (free software implementation based on the method of moments), for a specific case corresponding to Figure~\ref{fig_6}c,d: a rectangular array (lattice constants $a = 550$ nm and $b = 985$ nm) of semiconductor nanodisks of  diameter $D = 210$ nm and height $H = 200$ nm. The resulting reflectances are shown in Figure~\ref{fig_7} confirming that accidental BICs are observed in both polarizations with similar dispersion relations and Q factors. Nevertheless, there is a slight spectral/angular shift with respect to the predictions of our CEMD model, especially in TM polarization: the TE-BIC emerges at $\theta \sim 37^{\circ}$ and $\lambda \sim 902$ nm, whereas  the TM-BIC appears at $\theta \sim 41^{\circ}$ and $\lambda \sim 930$ nm. This is somewhat expected since our model only retains the dipolar responses. Still, the agreement is remarkable by all means (other resonant bands and related features are reproduced too), bearing in mind in turn that for high resolution reflectance maps, CEMD calculations take a few hours, about two orders of magnitude less than typical numerical simulations. Our CEMD can thus predict in a fast manner the range of parameters wherein double accidental BICs emerge in real configurations, to be fine tuned later on through full numerical simulations in a limited range of disk/lattice parameters.

Bear in mind that double accidental BICs can be very useful to enhance light-matter interactions within all-dielectric nanodisk metasurfaces: this implies that two resonances extending all over the metasurface with extremely large (in principle, diverging) Q-factors can be simultaneously excited with mixed polarized light, with the advantage that the arrays can be designed  to obtain ad-hoc  positions in the $(\omega,k)$ space (namely, wavelength and angle nearly at will). 

\section{Conclusions}

To summarize, with the help  of our coupled electric/magnetic dipole model for infinite planar arrays, valid for meta-atoms properly described by dipolar resonances, we have determined analytical conditions for the emergence of accidental BICs in all-dielectric metasurfaces. This is demonstrated first for all-dielectric spheres in rectangular arrays through explicit conditions that incorporate their dipolar responses through analytical Mie polarizabilities. We have explored the analytical conditions  in normalized dimensions that can  be easily extrapolated to any spectral regime as long as the refractive index of the spheres remains constant within the studied spectral domain; in particular, realistic nanosphere metasurfaces supporting accidental BICs in the optical domain can be inferred from them, using either oxides and semiconductors as sphere material. Moreover, such conditions are also exploited to design semiconductor metasurfaces supporting accidental BICs in the optical domain, typically fabricated through standard lithographic means, such as nanodisk arrays, by properly accounting for the polarizabilities of the cylinder-shaped meta-atoms. In this regard, thanks to the additional  degree of freedom (aspect ratio) allowed by the cylindrical shape as compared to spheres, we are able to determine conditions  for the emergence of, not only  single, but also double (both linear polarizations) accidental BICs at the same $\omega$ and $k_{||}$: this in turn implies that two resonances extending all over the metasurface with extremely large (in principle, diverging) Q-factors can be simultaneously excited with mixed polarized light, with the advantage that the arrays can be designed   to obtain ad-hoc  positions in the $(\omega,k)$ space (namely, wavelength and angle nearly at will). Such phenomenology opens a new venue for BIC-enhanced light-matter interaction in all-dielectric metasurfaces in the optical domain (and throughout the electromagnetic spectrum), paving the way to a variety of  phenomena that may benefit for the emergence of two accidental BICs away from the $\Gamma$-point for both polarizations, such as non-linear processes, lasing, chirality, etc. 



\medskip
\textbf{Supporting Information} \par 
Supporting Information is available from the Wiley Online Library or from the author. \par

\medskip
\textbf{Acknowledgements}\\
Financial support is acknowledged from the Swiss National Science Foundation through the project 197146 and from the Spanish Ministerio de Ciencia e Innovación (MICIU/ AEI/ FEDER, UE) through the grants PGC2018-095777-B-C21 (MELODIA) and  PID2019-109905GA-C22.
\par 

\medskip
\textbf{Conflict of Interest}\\
The authors declare no conflict of interest. \par 

\medskip
\textbf{Data Availability Statement}\\
The data that support the findings of this study are available from the corresponding author upon reasonable request. \par 

\medskip

%
\bibliographystyle{MSP}
\bibliography{library}





\begin{figure}
\textbf{Table of Contents}\\
\medskip
  \includegraphics[width=1\columnwidth]{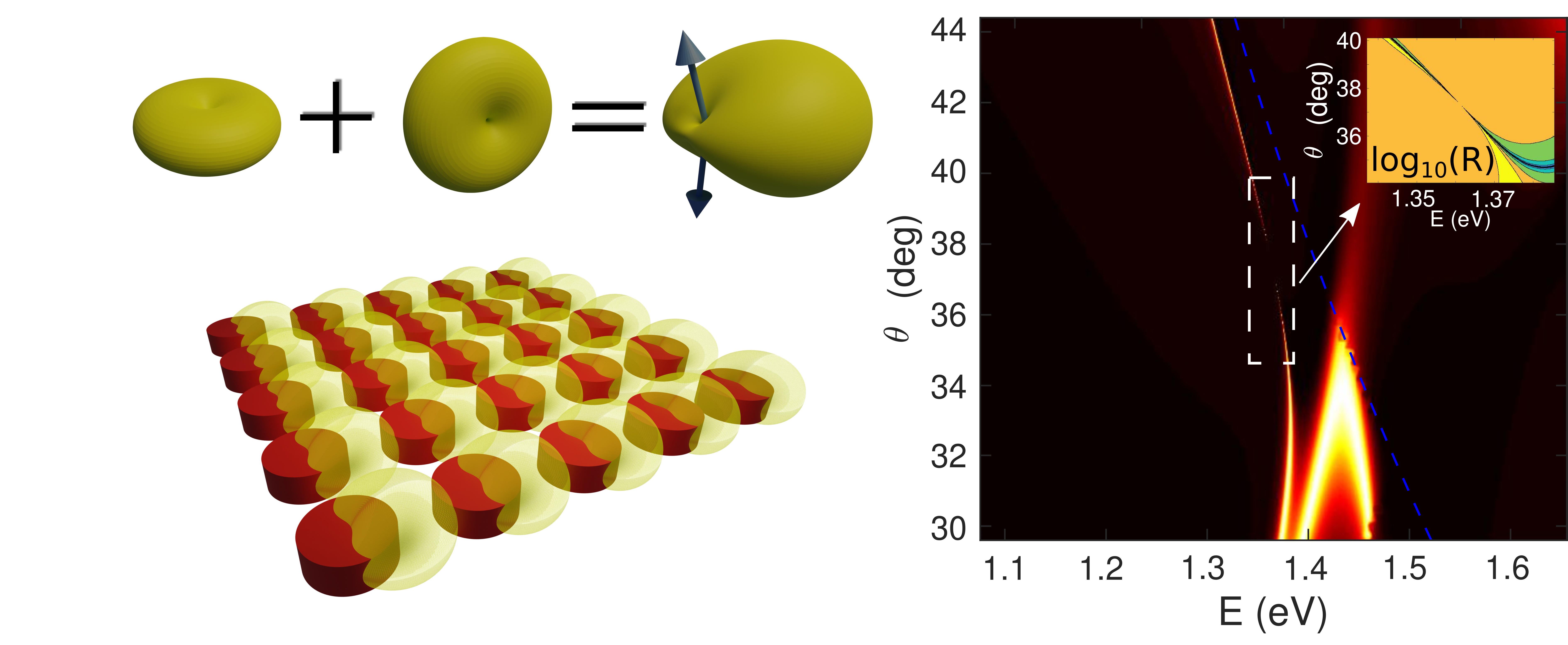}
  \medskip
  \caption*{Accidental bound state in the continuum sketch and reflectance map.}
\end{figure}

\end{document}